# A semi-automatic ultrasound image analysis system for the grading diagnosis of COVID-19 pneumonia

Yuanyuan Wang[1], Yao Zhang[2]*, Qiong He[1], Hongen Liao[1] and Jianwen Luo[1]*
[1]Department of Biomedical Engineering, School of Medicine, Tsinghua University, Beijing, China
[2]Department of Ultrasound, Beijing Ditan Hospital, Capital Medical University, Beijing, China
*Email: zgzsy007@163.com, luo_jianwen@tsinghua.edu.cn

*Abstract*—This paper proposes a semi-automatic system based on quantitative characterization of the specific image patterns in lung ultrasound (LUS) images, in order to assess the lung conditions of patients with COVID-19 pneumonia, as well as to differentiate between the severe / and no-severe cases. Specifically, four parameters are extracted from each LUS image, namely the thickness (TPL) and roughness (RPL) of the pleural line, and the accumulated with (AWBL) and acoustic coefficient (ACBL) of B lines. 27 patients are enrolled in this study, which are grouped into 13 moderate patients, 7 severe patients and 7 critical patients. Furthermore, the severe and critical patients are regarded as the severe cases, and the moderate patients are regarded as the non-severe cases. Biomarkers among different groups are compared. Each single biomarker and a classifier with all the biomarkers as input are utilized for the binary diagnosis of severe case and non-severe case, respectively. The classifier achieves the best classification performance among all the compared methods (area under the receiver operating characteristics curve = 0.93, sensitivity = 0.93, specificity = 0.85). The proposed image analysis system could be potentially applied to the grading and prognosis evaluation of patients with COVID-19 pneumonia.

## I. INTRODUCTION

Fast and accurate evaluation of lung conditions is significant for reducing the mortality of patients with coronavirus disease 2019 (COVID-19) pneumonia. Chest computed tomography (CT) is considered as a gold standard to evaluate the severity of pulmonary symptoms, but it is not universally available in the intensive care unit (ICU) to follow up the critically ill patients owing to the exposure to high-dose radiation and the frequent need of patient transport [1, 2].

Lung ultrasound (LUS) is currently used as a reliable imaging tool in the ICU to monitor the lung conditions of severe cases because it is portable, real-time and free from ionizing radiation [3, 4]. In a normal LUS image, the pleural line (PL) is a perfect reflector to the ultrasound pulse because of the acoustic impedance mismatch between the soft-tissues and the air inside the lung. The B lines (BLs) are comet tail artifact starting with the PL and extending to the edge of the screen. Specific image patterns of LUS are found to be associated with the severity of COVID-19 pneumonia in clinical diagnosis, including the lung consolidation, thickened PL, confluent BLs. These image patterns have been used in the qualitative or semi-quantitative diagnosis based on visual observations of the clinician [5, 6]. Various quantitative techniques have also been proposed by extracting parameters associated with the PL intensity or the BLs number, or using various deep learning networks for the grading diagnosis, and prognosis evaluation of patients diagnosed with COVID-19 pneumonia [7-12].

In this study, a semi-automatic automatic ultrasound image analysis system is proposed to assess the lung conditions of patients with COVID-19 pneumonia, through quantitative characterization of LUS images acquired from the moderate, severe and critical patients. The moderate patients are regarded as the non-severe cases. The severe and critical patients are regarded as the severe cases. Four biomarkers, namely the thickness (TPL) and roughness (RPL) of PL, as well as the accumulated width (AWBL) and the acoustic coefficient (ACBL) of BLs are extracted. The performances of these biomarkers in grading the severity of these patients and in the binary classification of the non-severe and severe cases are evaluated.

## II. METHODS

### A. Study protocol

This study was approved by the local ethics committee of Beijing Ditan Hospital. The need for written informed consent was waived due to the retrospective nature of the study. 27 patients diagnosed with COVID-19 pneumonia were included and grouped into 13 moderate patients, 7 severe patient and 7 critical patients following the Chinese clinical guidance for COVID-19 pneumonia diagnosis and treatment [13]. LUS examination was performed on each patient by an experienced clinician with a Hi Vision Preirus system (Hitachi Healthcare, Tokyo, Japan). The imaging depth and gain were set adaptively

This work was supported in part by Tsinghua University Spring Breeze Fund (2021Z99CFY025), the National Natural Science Foundation of China (61871251, 61801261, and 62027901), Sichuan Science and Technology Program (2019YFSY0048), Tsinghua-Peking Joint Center for Life Sciences, and the Young Elite Scientists Sponsorship by China Association for Science and Technology (Corresponding authors: Yao Zhang and Jianwen Luo).

Y. Wang and Y. Zhang* contributed equally to this work and were co-

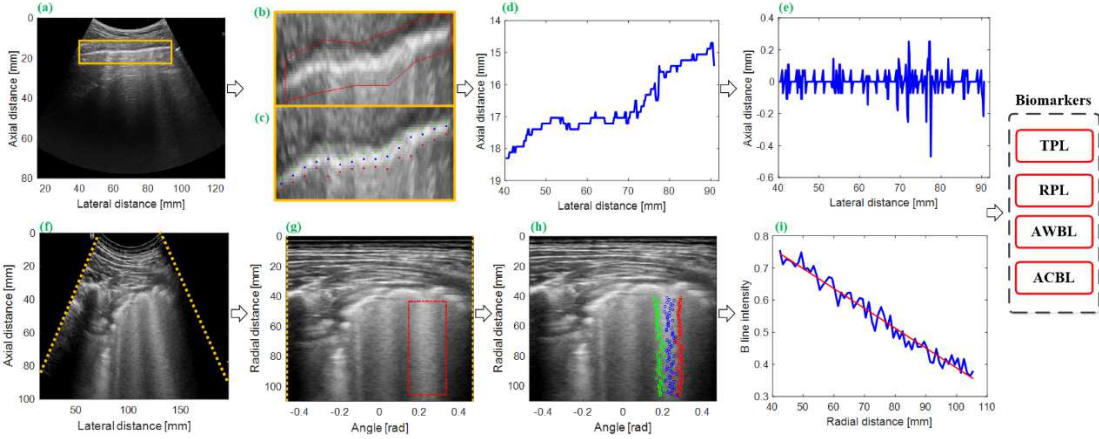

Fig. 1. The procedure of LUS image analysis. (a) An example of normalized LUS images for the illustration of PL analysis. (b) Manual selection of the PL. (c) Automatic detection of the anterior, central, and posterior locations of the PL. (d) The central depth profile of the PL. (e) The high-pass filtered profile of (d). (f) An example of normalized LUS images for the illustration of BL analysis. (g) Manual selection of the BL region from the transformed image obtained by mapping (f) into a linear grid through conversion from a polar to a Cartesian system. (h) Automatic detection of the left, central and right locations of the BL. (i) Linear progression for the calculation of ACBL.

to ensure the image quality suitable for clinical diagnosis. For each patient, a compete examination includes 12 scanning regions, including the anterior and superior lung region, anterior and inferior region, lateral and superior lung region, lateral and inferior lung region, posterior and superior lung region, and posterior and inferior lung region, for each half-chest.

### B. Quantitative characterization of LUS images

In the pre-processing steps, the LUS images were firstly normalized to minimize the influences of imaging depths and gains, respectively. Fig.1 illustrates the processing procedure for LUS image analysis. As shown, a rough outline of the PL, as indicated by the red lines in Fig. 1(b), was manually delineated. After that, the accurate central location of the PL, as indicated by the blue points was subsequently identified by detecting the peaks inside the image outline automatically. The anterior and posterior locations of the PL were detected by finding the line with 75% of the peak intensities and were indicated by the green and red points in Fig. 1(c)), respectively. The TPL was calculated as the mean difference between the depths of the posterior and anterior boundaries. The depth profile of the central PL was obtained (Fig.1 (d)) and a high-pass filter was used to eliminate the contour of the PL. The RPL was then calculated as the standard deviation (SD) of the filtered depth profile (Fig. 1(e)) to reflect the regularity and continuity of the PL.

For the analysis of BLs, we firstly map the LUS image (Fig. 2(f) into a linear grid through coordinate system conversation to obtain a transformed image. A rough outline was manually delineated around each BL, as indicated by the red rectangle in Fig. 2(g). The left (green points), central (blue points) and right locations (red points) of the BL inside the rectangle were automatically identified, with same detection method as that used for the PL. The mean distances between the left and right boundaries (normalized by the PL length) of different BLs were summed to obtain the AWBL. Linear regressions were performed between the peak intensities and their locations for different BLs, and the mean slope of different fitting lines was calculated as the ACBL.

### C. Statistical analysis

The least significance difference (LSD) tests based one-way analysis of variance (ANOVA) method was performed on the biomarkers of different groups, in order to assess the performance of the proposed method in evaluating the disease severity. A support vector machine (SVM) classifier was applied to the binary classification of non-severe case and severe case by combing all the extracted biomarkers as input. For each method, the classification performance was assessed by deriving the receiver operating characteristic (ROC) curve. The sensitivity (SEN), specificity (SPC) and area under the ROC curve (AUC) were also calculated.

### III. RESULTS

Figure 2 presents the comparison results of the PL- and BLs-related biomarkers for different groups, respectively. As shown, the biomarkers related to the PL cannot significantly differentiate between any two groups based on the LSD test results. Nevertheless, an increasing trend with the aggravation of the disease could be found for both the TPL and RPL. In terms of the BLs-related biomarkers, AWBL can significantly differentiate between the moderate patients and severe patients ($p < 0.01$), ae well as between the moderate and critical patients ($p < 0.001$), respectively. ACBL is found to differ significantly between the moderate and severe patients ($p < 0.05$), but the differences are not statistically significant between the critical and moderate patients, and between the critical and severe patients. An increasing trend could also be found for the AWBL and ACBL with the aggravation of the disease.

Figure 3 presents the ROC curves obtained using each biomarker and the SVM classifier with all the biomarkers as input, as well their respective AUC. The SEN and SPC obtained by finding the optimal cut-off value are also calculated. Compared with the TPL, the RPL shows improved quality of binary classification with increased SEN, SPC and AUC. ACBL and AWBL achieve the highest SPC and SEN for all of the biomarkers, respectively. Among all the compared methods, the SVM classifier achieves the best classification performance, with the highest AUC and SEN when all the extracted parameters are used as input.

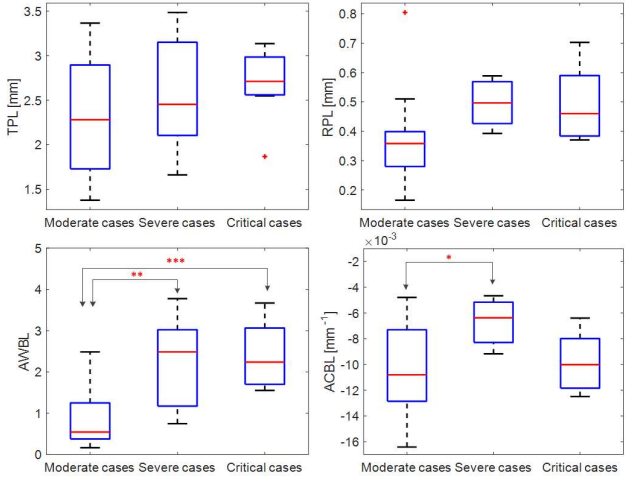

Fig. 2. Comparisons of the PL- and BLs-related biomarkers among different cases. *$p < 0.05$, **$p < 0.01$, ***$p < 0.001$.

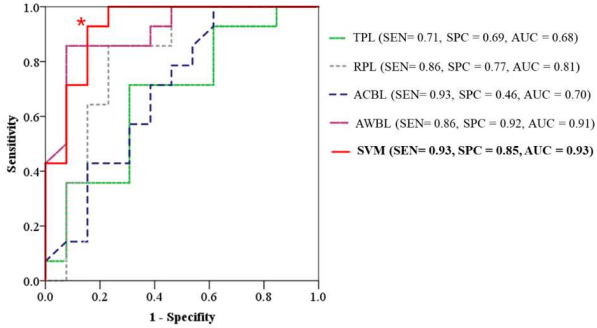

Fig. 3. Binary classification results of each biomarker and the SVM classifier with all the biomarkers as the input.

## IV. DISCUSSIONS

In clinical practice, severity assessment of COVID-19 pneumonia based on ultrasound images is conducted by visually identifying interested image patterns such as the thickened PL, confluent BLs, and presence of lung consolidations, or by using semi-quantitative scoring systems based on these patterns. However, such methods are subjective and rely much on the clinician's experiences. In this work, the PL- and BLs- related biomarkers are extracted through quantitative analysis of LUS B-mode images to evaluate the severity of lung condition.

In previous studies, various methods have been proposed to quantitatively assess the severity of lung condition. These methods mainly focused on the automatic detection of BLs o and PL, and extracted parameters related to number of BLs and pixel intensity of the PL. However, the properties of BLs and PL are characterized separately and are not combined for the grading diagnosis [11, 12]. Some artificial intelligence based methods have achieved satisfactory performances in the prediction and grading of the disease severity, as well as the prognosis evaluation of lung conditions. These methods usually use the quantitative parameters extracted with conventional technique, the ultrasound images/videos, the clinical information, or the combinations of these features as the input of the networks [7-12]. However, the disadvantages are poor interpretability and the requirement of large number of annotated samples.

In this study, we proposed a semi-automatic ultrasound image analysis system for the disease severity assessment and binary classification of the non-severe and severe patients, by extracting the biomarkers characterizing the ultrasound patterns that the clinicians are interested in.

TPL and RPL are extracted for the PL analysis, which reflect the thickening and irregularity degree of the PL, respectively. The TPL and RPL are found to increase with the aggravation of the disease severity, which are consistent with the clinical observations. AWBL and ACBL are extracted for the BL analysis, which reflect the confluence and attenuation degree the BLs. For the analysis of BLs, a larger AWBL is an indicator of the confluence of BLs. As is found, AWBL and ACBL differ significantly between the moderate patients and severe patients, but the differences between the severe patients and critical patients were not statistically significant using any biomarker, which may be because the classification standard for different cases is not directly related to the lung condition, but reflect the comprehensive clinical state of the patients. Therefore, the severe and critical patients may have similar lung conditions.

The binary classification between the severe and non-severe patients is important in the clinical diagnosis for timely treatment of the critically ill patients and effective reduction of the overall mortality of COVID-19 pneumonia. The SVM classifier with all the biomarkers as the input achieves better classification performance than any proposed biomarker taking the advantage of more comprehensive information. In the future, more biomarkers related to the PL and BLs will be dug out to further improve the classification performance, such as the mean and standard deviation of the PL intensities, the intensity of the BLs, and the statistical analysis of the BLs speckles.

Lung consolidation is an important indicator of the disease severity, but has not been analyzed in this study. Because lung consolidation only exists in specific patients and thus it is hard to compare the conditions of different patients with and without the presence of lung consolidation using related indices fairly.

The rough outlines of the PLs and BLs need to be manually delineated before accurate identification of their central positions, leading to a semi-automatic nature of the proposed method. In addition, different cutoff thresholds used for the determination of the BL and PL boundaries are tested in our preliminary experiment and a cutoff threshold of 75% is finally selected, which provides the boundaries closest to those determined by the clinicians. In the future work, the central locations and boundaries of PL / BL may be automatically detected using the image segmentation algorithms and deep learning networks. Ultrasound images from patients at different stages of treatment could be collected to investigate the capability of the proposed method in the prognosis evaluation.

In our previous work [14], biomarkers including the mean and standard deviation of the PL, and the accumulated intensity of BL were extracted and combined as the input of the classifier. However, Some of the parameters are closely associated with the image intensity, which will be affected by the varying imaging parameters such as TGC and focal depth, and are thus sensitive to the imaging parameters. In contrast, all the biomarkers in this study are hardly affected by those parameters, and thus achieve improved robustness, as well as keep comparable binary classification performance at the same time.

## V. Conclusions

In this study, we propose a semi-quantitative analysis method for assessing the severity of COVID-19 pneumonia by characterizing imaging patterns related to PLs and BLs. Among all the extracted biomarkers, AWBL differs significantly between the moderate and severe patients, as well as between the moderate and critical patients. The SVM classifier with all the biomarkers as the input show the optimal binary classification performance between the severe and non-severe cases (SEN = 0.93, SPC = 0.85, AUC = 0.93). The proposed method may be potentially applied to the automatic grading and prognostic evaluation of patients with COVID-19 pneumonia.